\documentclass[aps,11pt,tightenlines,nofootinbib]{revtex4}
\usepackage{amsmath,amscd,amssymb}
\usepackage{color,epsfig,ulem}

\def\beq{\begin{equation}}
\def\eeq{\end{equation}}
\def\be{\begin{eqnarray}}
\def\ee{\end{eqnarray}}

\begin{document}

\title{Gauge Invariance and Vacuum Energies of Non--Abelian
String--Configurations}

\author{H. Weigel$^{a)}$ and M. Quandt$^{b)}$}

\affiliation{
$^{a)}$Physics Department, Stellenbosch University,
Matieland 7602, South Africa\\
$^{b)}$Institute for Theoretical Physics, T\"ubingen University,
D--72076 T\"ubingen, Germany}

\begin{abstract}
We improve on a method to compute the fermion contribution to the
vacuum polarization energy of string--like configurations in a
non--Abelian gauge theory. We establish the new
method by numerically verifying the invariance under (a subset of)
local gauge transformations. This also provides further support for
the use of spectral methods to compute vacuum polarization energies
in general. We confirm that the vacuum energy in the
$\overline{\rm MS}$ renormalization scheme is tiny as compared to the
mass of the fluctuating fermion field.
Numerical results for the physical {\it on--shell}
scheme are also presented.
\end{abstract}

\maketitle

\section{Introduction}

\noindent
The electro--weak  sector of the standard model suggests the existence of extended 
flux-tubes called $Z$-- or cosmic strings, which may have profound consequences for 
cosmological
questions~\cite{Vilenkin:1984ib,Brandenberger:1998rc,Achucarro:2008fn,Copeland:2009ga}.
These configurations are, however, not protected by any topological argument and thus
are classically unstable~\cite{Achucarro:1999it,Nambu:1977ag}. To investigate 
whether quantum effects provide dynamical stabilization, it is very important 
to compute the vacuum polarization energy $\Delta E$ of the string configuration
{\it\,reliably}.  In ref.~\cite{Weigel:2009wi} we have recently provided
a proof--of--principle computation of $\Delta E$ in an $SU(2)_L$ gauge
theory in three spatial dimensions\footnote{See also 
ref.~\cite{Weigel:2009wi} for a discussion of earlier 
attempts~\cite{Bordag:1998tg,Bordag:2003at,Langfeld:2002vy,Graham:2004jb,AlonsoIzquierdo:2004ru,Graham:2006qt,Schroder:2007xk,Groves:1999ks,Bordag:2002sa,Baacke:2008sq,Davis:1999ec}
to estimate $\Delta E$ for string configurations, both in Abelian and non--Abelian 
theories.}. That approach had the drawback that it required
the introduction of an auxiliary field at spatial infinity to make various
components of the calculation well--defined.  In the present letter we
demonstrate that the formulation simplifies considerably in a suitable set of
gauges. In particular, the use of an auxiliary field at infinity is
avoided altogether.

The string configuration is translationally invariant along its symmetry axis
(which we choose to be $\hat{z}$), {\it i.e.}~it only depends on the
distance $\rho$ from the axis and the corresponding azimuthal angle~$\varphi$.
Finiteness of the classical energy (per unit length) requires that the 
string configuration must be pure gauge at spatial infinity, which turns out 
to have a non--trivial angular dependence due to the winding of the string.
As a consequence, gauge variant functionals of the string fields, such as Feynman
diagrams, are ill--defined. This is the major obstacle for a straightforward 
application of the spectral methods~\cite{Graham:2009zz} to compute the vacuum 
polarization energy of a string. In ref.~\cite{Weigel:2009wi} this obstacle was 
circumvented by the introduction of a {\it return string} that unwound the 
fields at spatial infinity. In numerical calculations this has the disadvantage 
that spatial infinity can only be reached by extrapolation to very extended 
profiles of the return string. These wide extensions induce large impact 
parameters so that channels with very large angular momenta must be considered.
Here we argue that there are particular gauges in which the computation of
$\Delta E$ does not require any return string. The litmus test then is to
establish the invariance of $\Delta E$ under changes of parameters that
classify these gauges. We present numerical evidence to confirm that this
is indeed the case. Our finding gives further support for the use of spectral 
methods in general, as it proves the equality of gauge variant and divergent 
Feynman diagrams, and (equally gauge variant and divergent) terms in the 
Born series.

In more detail, the string configuration consists of $SU(2)$ vector and Higgs 
fields, $W^\mu$ and $\Phi$, respectively:
\begin{eqnarray}
 \vec{W}&=&n\,{\rm sin}(\xi_1)\,\frac{f_G(\rho)}{g\rho}\,\hat{\varphi}\,
\begin{pmatrix}
{\rm sin}(\xi_1) & i {\rm cos}(\xi_1)\,{\rm e}^{-in\varphi} \\[2mm]
-i {\rm cos}(\xi_1)\,{\rm e}^{in\varphi} & - {\rm sin}(\xi_1)
\end{pmatrix}
\qquad {\rm and} \quad \nonumber \\[4mm]
\Phi&=&v \, f_H(\rho)
\begin{pmatrix}
{\rm sin}(\xi_1)\, {\rm e}^{-in\varphi} & -i {\rm cos}(\xi_1) \\[2mm]
-i {\rm cos}(\xi_1) & {\rm sin}(\xi_1)\, {\rm e}^{in\varphi}
\end{pmatrix}\,.
\label{string}
\end{eqnarray}
Here, we have used temporal gauge $W^0 = 0$ and the $SU(2)$ isospin structure 
is written in explicit matrix notation. Moreover, $v$ is the (classical)
vacuum expectation value of the Higgs field and $g$ is the gauge coupling 
constant introduced in the string parameterization for later convenience.

The configuration~(\ref{string}) is commonly called a Z--string, because the
corresponding component $Z\sim {\rm tr}_I(W\tau_3)$ exhibits the spatial
dependence of an Abelian string. The radial functions $f_G(\rho)$ and $f_H(\rho)$
approach unity at spatial infinity while they vanish for $\rho=0$. They are the
typical profiles of the Nielson--Olesen type of string~\cite{Nielsen:1973cs}.
The angle $\xi_1 \in [0,\pi]$ is a free parameter that determines the
relative weight of the gauge and Higgs profiles; it also measures the
fractional flux carried by the Z--string.

We are mainly interested in the contribution from the fermion fluctuations to
the vacuum polarization energy because it dominates the boson contribution when
the number $N$ of other internal degrees of freedom ({\it e.g.}~color) becomes
large. Motivated by the standard model we consider a non--Abelian gauge theory
in which the gauge field only couples to left--handed fermions.
The fermion--string interaction is then given by the Lagrangian
\begin{equation}
\mathcal{L}_\Psi=\overline{\Psi}i\gamma_\mu
\left(\partial^\mu-igW^\mu\right) P_L \Psi
+\overline{\Psi}i\gamma_\mu \partial^\mu P_R \Psi
-f\,\overline{\Psi}\left(\Phi P_R+\Phi^\dagger P_L\right)\Psi\,,
\label{gaugelag}
\end{equation}
where $P_{R,L}=\frac{1}{2}\left(1\pm\gamma_5\right)$ are projection
operators on right-- and left--handed components, respectively. The
strength of the Higgs--fermion interaction is parameterized by the Yukawa
coupling constant $f$, so that the fermions acquire the mass $m = vf$
via spontaneous symmetry breaking.

\section{Calculational Techniques}

\noindent
We extract the Dirac Hamiltonian from eq.~(\ref{gaugelag}) and perform
a local gauge transformation  $H\to U^\dagger H U$ where
\begin{equation}
U=-iP_L\tau_1{\rm exp}\left(i\,\hat{n}\cdot\vec{\tau}\,\xi\right)+P_R
\qquad {\rm with}\qquad
\hat{n}=\begin{pmatrix}
{\rm cos}(n\varphi) \cr -{\rm sin}(n\varphi) \cr 0
\end{pmatrix} \,.
\label{gaugetf}
\end{equation}
Here $\xi=\xi(\rho)$ is an arbitrary radial function that defines a subset
of gauge transformations. The transformed Dirac Hamiltonian becomes
\begin{eqnarray}
H&=&-i\begin{pmatrix}0 & \vec{\sigma}\cdot\hat{\rho} \cr
\vec{\sigma}\cdot\hat{\rho} & 0\end{pmatrix} \partial_\rho
-\frac{i}{\rho}\begin{pmatrix}0 & \vec{\sigma}\cdot\hat{\varphi} \cr
\vec{\sigma}\cdot\hat{\varphi} & 0\end{pmatrix} \partial_\varphi
+m\begin{pmatrix} 1 & 0 \cr 0 &-1\end{pmatrix}
+H_{\rm int}\,, \label{eqDirac0}\\[4mm]
H_{\rm int}&=&
m\left[\left(f_H{\rm cos}(\delta\xi)-1\right)
\begin{pmatrix} 1 & 0 \cr 0 &-1\end{pmatrix}
+if_H\,{\rm sin}(\delta\xi)\begin{pmatrix}0 & 1 \cr -1 & 0\end{pmatrix}
\hat{n}\cdot\vec{\tau}\right]
+\frac{1}{2}\frac{\partial \xi}{\partial \rho}
\begin{pmatrix}-\vec{\sigma}\cdot\hat{\rho}
& \vec{\sigma}\cdot\hat{\rho} \cr
\vec{\sigma}\cdot\hat{\rho}
& -\vec{\sigma}\cdot\hat{\rho}\end{pmatrix}\hat{n}\cdot\vec{\tau}
\nonumber \\[3mm]
&&
+\frac{n}{2\rho}\, \begin{pmatrix}-\vec{\sigma}\cdot\hat{\varphi}
& \vec{\sigma}\cdot\hat{\varphi} \cr
\vec{\sigma}\cdot\hat{\varphi}
& -\vec{\sigma}\cdot\hat{\varphi}\end{pmatrix}
\Big[f_G\,{\rm sin}(\delta\xi)\,I_G(\delta\xi)
+(f_G-1)\,{\rm sin}(\xi)\,I_G(-\xi)\Big]\,.
\label{eqDirac}
\end{eqnarray}
We have made explicit the dependence on the angles $\delta\xi\equiv\xi_1-\xi$
and $\xi$ via the isospin matrix
\begin{equation}
I_G(x)=\begin{pmatrix}-{\rm sin}(x) & -i{\rm cos}(x)\,{\rm e}^{in\varphi} \\[2mm]
i{\rm cos}(x)\,{\rm e}^{-in\varphi} & {\rm sin}(x) \end{pmatrix} \,,
\label{defIG}
\end{equation}
while the explicit matrices in eqs.~(\ref{eqDirac0}) and (\ref{eqDirac})
act in spinor space.

The key idea is now to impose the boundary conditions $\xi(0)=0$ and
$\xi(\infty)=\xi_1$. Together with the boundary conditions for the physical
profiles $f_G$ and $f_H$ this defines a well--behaved scattering problem
for which a scattering matrix and, more generally, a Jost function for
momenta in the upper half complex plane can be straightforwardly computed.
Furthermore, the Born series to these scattering data can be constructed
by iterating $H_{\rm int}$. In contrast to the exact Jost function, the
individual terms in this series are gauge dependent, {\it i.e.}~they
vary with $\xi(\rho)$. After collecting these ingredients we proceed as in
ref.~\cite{Weigel:2009wi}:

\begin{enumerate}
\item 
In each angular momentum channel we evaluate the Jost function for imaginary 
momenta from the Dirac Hamiltonian, eq.~(\ref{eqDirac}). To this end we 
continue the momentum $k$, that is conjugate to the radial coordinate $\rho$, 
analytically by substituting $k\to i\tau$ with $\tau$ being a real variable. 
{}From the Jost function we then subtract its first and second orders of the 
corresponding Born series. This difference is summed over angular momenta. 
The analytic continuation to imaginary momenta is important because it allows 
us to exchange sums over angular momenta with momentum 
integrals~\cite{Schroder:2007xk} and implicitly
accounts for the bound state contribution to $\Delta E$.

\item
We introduce a {\it fake boson} field whose second Born
order approximation of the Jost function has the same divergence structure as
the combined third and fourth order Born terms for the fermion problem.
Again this quantity is summed over angular momenta. The fake boson method is 
a computational trick to circumvent the very cumbersome evaluation of third and 
fourth order Born terms and Feynman diagrams. This simplification has been
established for purely logarithmic divergent contributions.
\item
We integrate the difference of the two functions
constructed above over imaginary momenta~$\tau$, weighted by a kinematical
factor characteristic for string--like configurations that are translationally
invariant along a fixed direction~\cite{Graham:2001dy}. The value of this
integral is the phase shift contribution, $\Delta E_\delta$.
\item
We add back the first and second order Born contributions in
form of renormalized Feynman diagrams of identical order in $H_{\rm int}$.
We call this piece $\Delta E_{\rm FD}$ and discuss the details
of the necessary counterterms further below.
\item
Finally, we add back $\Delta E_{\rm B}$ which is the renormalized
second order fake boson Feynman diagram that corresponds to the
subtraction under 2. It should be emphasized that the renormalization of
$\Delta E_{\rm B}$ is accomplished by the counterterms
in the fermion sector.
\end{enumerate}

In total, the fermion contribution to the renormalized vacuum polarization
energy per unit length of the string reads
\begin{equation}
\Delta E= \Delta E_\delta+\Delta E_{\rm FD}+\Delta E_{\rm B}\,.
\label{finalresult}
\end{equation}

\section{Numerical Results}

\noindent
In addition to the angle $\xi_1$, the parameterization of the string background 
with the above motivated boundary conditions introduces three  width parameters, 
$w_H$, $w_G$ and $w_\xi$,
\begin{equation}
f_H(\rho)=1-{\rm e}^{-\frac{\rho}{w_H}}\,,\qquad
f_G(\rho)=1-{\rm e}^{-\left(\frac{\rho}{w_G}\right)^2}
\qquad {\rm and} \qquad
\xi(\rho)=\xi_1\left[1-
{\rm e}^{-\left(\frac{\rho}{w_\xi}\right)^2}\right]\,.
\label{fields}
\end{equation}
This parameterization guarantees that the interaction Hamiltonian,
$H_{\rm int}$ is well defined at $\rho\to0$ and no $1/\rho$ type
singularity is encountered. Obviously, the litmus test for our
calculation is that the final result for $\Delta E$ must not depend on the scale
$w_\xi$ introduced in the gauge transformation profile. In our numerical
studies we always assume the special case $n=1$.

Numerically the most cumbersome quantity is the phase shift contribution
$\Delta E_\delta$. For small values of the scale parameters $w_H$ and $w_G$,
in particular, the calculation of $\Delta E_\delta$ for a single background
configuration takes several days of CPU time on a modern desktop computer.
In the treatment of ref.~\cite{Weigel:2009wi} at least as much time is consumed 
for {\it each} set of variational parameters that characterizes the auxiliary 
return string.\footnote{ The calculation in ref.~\cite{Weigel:2009wi} needs
to be redone several times with varying sets of return string parameters
in order to extrapolate to an infinitely distant return string.}
Technically, we compute the momentum integral in $\Delta E_\delta$ with the
methods described above only
up to a numerical cut--off $\tau_{\rm max}$. For $\tau > \tau_{\rm max}$, we
approximate the integrand by an inverse power--law. The numerical cost of this
approach is determined by the smallest width parameter in
the problem: the smaller this width, the larger we have to take
$\tau_{\rm max}$ for the power--law approximation to be accurate.
Since a larger value for $\tau_{\rm max}$ also entails that more angular
momentum channels must be summed, the numerics become quickly expensive
for small widths.
{}From various integration methods and treatments of the contributions from
large angular and linear momenta, we estimate an overall numerical
accuracy of 1--2 \%, where the upper limit mainly applies to small widths.

\subsection{Verification of the method}

Before turning to the full string problem we  note
that the fake boson simplification introduces additional parameters
into the numerical calculation. We have numerically verified that
these parameters have no effect on the final result.

To verify the method, we will establish the invariance of the
vacuum polarization energy within the subset of gauge transformations
obtained by varying the width $w_\xi$ of the gauge transformation
profile $\xi(\rho)$. It is sufficient to consider the
$\overline{\rm MS}$ renormalization scheme because any other scheme
differs by finite counterterms that are manifestly gauge invariant
functionals of the background field.
We augment the $\overline{\rm MS}$ scheme by the no--tadpole condition
which adds the counterterm
\begin{equation}
\mathcal{L}_3=\frac{c_3}{2}\,{\rm tr}_I\left[\Phi^\dagger \Phi-v^2\right]
\label{lct3}
\end{equation}
such that the local first order Feynman diagram is exactly canceled.
The corresponding result is shown in table~\ref{tab_wxi}
for a typical set of parameters.
\begin{table}[t]
\centerline{
\begin{tabular}{c|ccc|c}
$w_\xi$ & \rule{2mm}{0mm}$\Delta E_{\rm FD}$\rule{2mm}{0mm}
& \rule{2mm}{0mm}$\Delta E_\delta$\rule{2mm}{0mm}
&\rule{2mm}{0mm} $\Delta E_{\rm B}$\rule{2mm}{0mm}
& \rule{2mm}{0mm}$\Delta E$ \rule{2mm}{0mm} \\[2mm]
\hline
0.5 \rule[-0mm]{0mm}{5mm} & -0.2515 & 0.3489 & 0.0046 & 0.1020 \\[2mm]
1.0 \rule[-2mm]{0mm}{5mm} & -0.0655 & 0.1606 & 0.0032 & 0.0983 \\[2mm]
2.0 \rule[-2mm]{0mm}{5mm} & -0.0358 & 0.1294 & 0.0038 & 0.0974 \\[2mm]
3.0 \rule[-2mm]{0mm}{5mm} & -0.0320 & 0.1235 & 0.0056 & 0.0971 \\[2mm]
4.0 \rule[-2mm]{0mm}{5mm} & -0.0302 & 0.1193 & 0.0080 & 0.0971
\end{tabular}}
\centerline{\parbox[l]{12cm}{\caption{\label{tab_wxi}
Independence on the scale of the gauge transformation parameter.
The other parameters are $w_H=w_G=2$
and $\xi_1=0.4\pi$, {\it i.e.}~the gauge field is fairly strong.
All energies are given in units of the classical fermion mass $m = vf$.}}}
\end{table}
Since we measure all energies in units of the fermion mass $m=vf$ and
all lengths in its inverse, in the $\overline{\rm MS}$ scheme
$\Delta E$ only depends on the specific shape of the background profiles.

We observe that the variation of the total result, $\Delta E$,
with $w_\xi$ is significantly less than the estimated numerical
error for $\Delta E_\delta$, even though some components of $\Delta E$
change by almost an order of magnitude in the considered
range\footnote{Considering even smaller values for $w_\xi$ becomes
numerically even more expensive
because the asymptotic behavior of the integrand for $\Delta E_\delta$
sets in at momenta roughly proportional to $1/w_\xi$.} of $w_\xi$.
We find similar results for other variational parameters $w_H$, $w_G$ and 
$\xi_1$. Within the numerical precision this confirms the gauge invariance 
of the vacuum polarization energy, at least for the subset
of gauge transformation that we have tested. 

We have also verified that $\xi_1\to\pi-\xi_1$ leaves $\Delta E$ 
unchanged within the numerical precision. This symmetry follows from the 
fact that for the fields, eqs.~(\ref{string}) this transformation equals a 
rotation by $\pi$ about the $\hat{z}$--axis in isospace. However, acting 
with this rotation on the gauge transformation $U$ in eq.~(\ref{gaugetf}) 
gives a completely different radial function $\xi(\rho)$ with appropriately
modified boundary values. The exact Jost function remains unchanged while 
neither the Born terms nor the Feynman diagrams are separately invariant, 
only their combination is.

It should be emphasized again that in the course of this computation, we have
added and subtracted formally identical quantities that are {\it per se} divergent 
and gauge variant. Hence our study also confirms that their finite pieces are 
identical, an assertion that is vital for the use of our spectral methods.
Previously, such an identity had only be shown for the leading order of the
Born and Feynman series within dimensional regularization~\cite{Graham:2009zz}.

The numerical data also confirms our previous result~\cite{Weigel:2009wi} that
$\Delta E$ is very small (as compared {\it e.g.}~against the
the fermion mass $m$) within the $\overline{\rm MS}$ scheme. Our previous
findings were, however, less accurate since they also required an
extrapolation to an infinitely distant return string.

In the left panel of figure~\ref{fig1} we display the dependence
of $\Delta E$ (in the $\overline{\rm MS}$ scheme) on the angle $\xi_1$
that characterizes the relative strength of the gauge field and Higgs background.
While $\Delta E(\xi_1)$ is monotonously increasing with $\xi_1$
({\it i.e.}~with stronger gauge fields) it develops a minimum around
$\xi_1=\pi/4$ when the width of the gauge field background is small.

\subsection{On--shell renormalization}

\noindent
To discuss physical implications we need to impose the on--shell
renormalization scheme. For an $SU(2)_L$ gauge theory the on--shell
renormalization conditions and the corresponding determination
of the counterterm coefficients have been discussed in
ref.~\cite{Farhi:2003iu}.\footnote{
Though the renormalized Feynman diagram is properly displayed
in ref.~\cite{Farhi:2003iu}, the formula for $c_4$ is missing
an overall factor $1/2$.}
To pass from $\overline{\mathrm{MS}}$ to on--shell, we have to add
the finite and manifestly gauge invariant counterterms
\begin{equation}
\mathcal{L}_{\rm ct}=
c_1{\rm tr}_I \left[W_{\mu\nu}W^{\mu\nu}\right]
+\frac{c_2}{2}\,{\rm tr}_I\left[
\left(\left(\partial_\mu-igW_\mu\right)\Phi\right)^\dagger
\left(\partial^\mu-igW^\mu\right)\Phi\right]
+\frac{c_4}{4}\left({\rm tr}_I\left[\Phi^\dagger \Phi-v^2\right]\right)^2\,,
\label{lct}
\end{equation}
where $W_{\mu\nu}=\partial_{[\mu}W_{\nu]}-ig[W_\mu,W_\nu]$
is the field strength tensor.

\begin{figure}
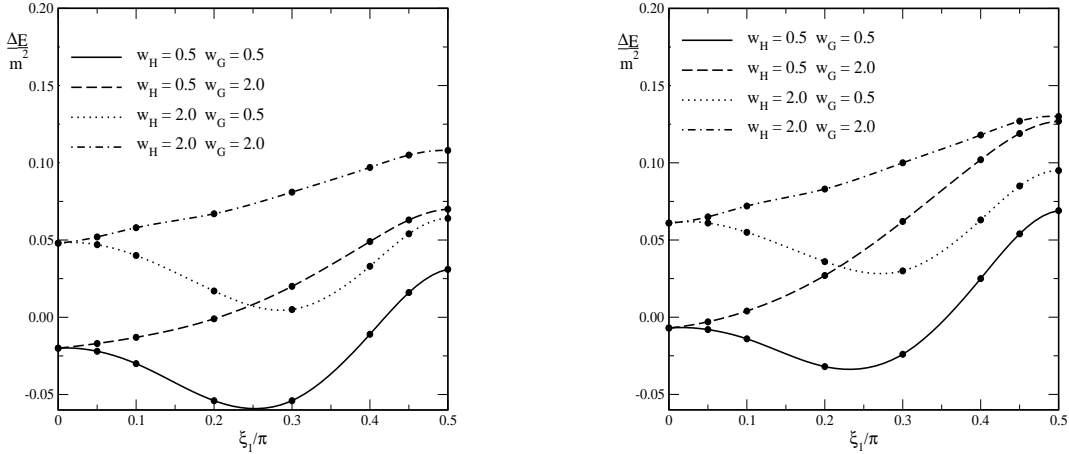

\centerline{
\epsfig{figure=msbar.eps,height=6.0cm,width=6.0cm}\hspace{2cm}
\epsfig{figure=onshell.eps,height=6.0cm,width=6.0cm}}
\smallskip\smallskip
\caption{\label{fig1}The parameter dependence of the vacuum polarization 
energy. Left panel: $\overline{\rm MS}$ scheme, right panel: on--shell scheme.
The dots denote the computed data and lines are spline interpolations.}
\end{figure}

The on--shell renormalization condition implies that the pole of the Higgs
propagator remains at the tree level mass, $m_h=m_h^{(0)}$, with unit residue.
This fixes the coefficients $c_2$ and $c_4$ and ensures the usual one--particle
interpretation of the states created by the asymptotic Higgs field.
Furthermore, we also demand that the residue of the gauge field propagator
(in unitary gauge) is unity, so that asymptotic $W$--fields create
one--particle $W$--boson states. This condition determines~$c_1$.
The position of the pole in the gauge boson propagator is then a prediction,
{\it i.e.}~the physical (on--shell) $W$-boson mass receives radiative corrections.
In our conventions (with all energies measured in units of $m=fv$)
we have $f=1/v$, so that $f^2=2\sqrt{2}m^2 G_F$ makes contact 
to the standard model parameters. Using $f=0.9$ and $g=0.7$ approximately
reproduces the top--quark and $W$--boson masses. Furthermore we use
$\mu_h=m_h/m=v/\sqrt{2}\approx0.8$. The corresponding results for
$\Delta E$ are shown in the right panel of figure~\ref{fig1}.
The additional counterterm contribution in the on--shell scheme
increases $\Delta E$ slightly making its little binding
effect from the $\overline{\rm MS}$ scheme even smaller.

\subsection{Total Energy}

\begin{figure}
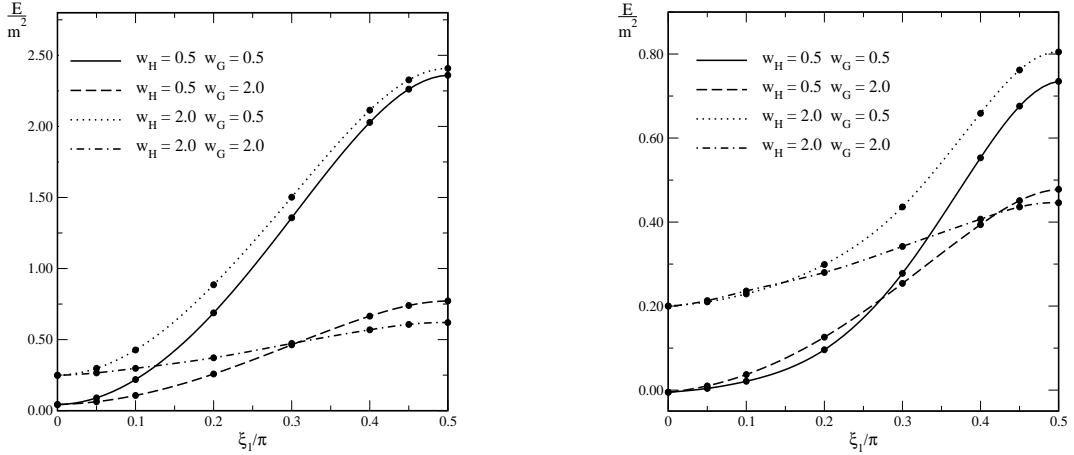

\centerline{
\epsfig{figure=total5.eps,height=6.0cm,width=6.0cm}\hspace{2cm}
\epsfig{figure=total10.eps,height=6.0cm,width=6.0cm}}
\smallskip\smallskip
\caption{\label{fig2}Total energy, eq.~(\ref{etot}) for $N=3$, in the on--shell 
scheme. Left panel: $g=f=5.0$, right panel: $g=f=10.0$. Note the difference
in the scales on the vertical axes.}
\end{figure}

\noindent
So far we have not considered the leading contribution to the
energy per unit length of the string, {\it i.e.}~the
classical energy~\cite{Graham:2006qt}
\begin{equation}
\frac{E_{\rm cl}}{m^2}=2\pi\int_0^\infty \rho d\rho\,\left\{
n^2\sin^2 \xi_1\,\biggl[\frac{2}{g^2}
\left(\frac{f_G^\prime}{\rho}\right)^2
+\frac{f_H^2}{f^2\rho^2}\,\left(1-f_G\right)^2\biggr]
+\frac{f_H^{\prime2}}{f^2}
+\frac{\mu_h^2}{4f^2}\left(1-f_H^2\right)^2\right\}\,,
\label{eq:ecl}
\end{equation}
where all quantities under the integral are dimensionless. 
Assuming that there are $N$ internal degrees of freedom, {\it e.g.}
$N=3$ for color, the total energy 
\begin{equation}
E=E_{\rm cl}+N\Delta E
\label{etot}
\end{equation}
will always be larger than $N\Delta E$
at the present level of approximation since $E_{\rm cl}$ is positive definite.
In order to allow for quantum stabilization, the classical energy must be
comparable to $\Delta E$, {\it i.e.}~tiny as well. This is not the case for 
standard model motivated parameters, which give $E_{cl}\sim 10\,{\rm m^2}$. 
Eq.~(\ref{eq:ecl}) shows that small a $E_{\rm cl}$ requires large coupling 
constants $g$ and $f$, or equivalently large masses of the fluctuating fermion.
To demonstrate this behavior we consider $E_{\rm cl}+3\Delta E$  for $g=f=5.0$ 
and $g=f=10.0$ in figure~\ref{fig2}.
For fermion masses of order $1.5\,{\rm TeV}$
we indeed observe a small binding as $E<0$ for narrow fields.
However, this may merely reflect the onset of the Landau
ghost~\cite{Ripka:1987ne,Hartmann:1994ai}. It thus seems unlikely that the
$Z$--string can be stabilized by fluctuating fermions without adding
fermion charge.

\section{Conclusion}

\noindent
We have presented a reliable computation of the vacuum polarization
energy that originates from fermion fluctuations about a cosmic string.
The present method is significantly more efficient than the only one
available so far~\cite{Weigel:2009wi}, because it makes redundant
the introduction of an auxiliary field near spatial infinity.
We have resolved the obstacles that stem from the non--trivial
structure of the individual string fields at spatial infinity
by choosing a subset of gauges for which the scattering problem
is well--behaved. We have verified the novel method by establishing
invariance with respect to gauge transformations within this subset. This
is far from trivial because in the process of computation formally
identical but divergent gauge variant quantities are added and
subtracted. As an important side--product we have generated further
support for the approach to compute vacuum polarization contributions to 
observables by spectral methods~\cite{Graham:2009zz}.

Our extensive numerical investigations indicate that the fermion
contribution to the vacuum polarization energy produces some binding
but it is far too small to overcome the large classical energy and
fully stabilize cosmic strings, at least for parameters that are
motived from the standard model.

Another stabilization scenario has been suggested in the $D=2+1$
model of ref.~\cite{Graham:2006qt}. Due to symmetry restoration
in the core of the string, the Higgs condensate vanishes locally and 
a significant number of bound states can be induced.  Population of 
these bound states may generate a charged object that is energetically  
favored against an equal number of free fermions with mass $m=vf$. 
We stress that this binding energy is of the same order in the
$\hbar$--expansion as the part of the vacuum polarization energy that we have
computed here. Hence the present calculation is a necessary ingredient in 
a future study of the quantum stabilization of charged cosmic strings.
This study will be subject of a future paper that will also serve to 
provide the details of the present computation.

\section*{Acknowledgments}
\noindent
We thank N. Graham and O. Schr\"oder for valuable discussion in
early stages of this project.

\end{document}